\documentclass[article,showpacs,10pt]{revtex4}
\usepackage{amssymb}
\usepackage{epsfig}
\usepackage{graphicx}
\usepackage{bm}

\begin{document}

\title{Towards a model for protein production rates}
\author{J. J. Dong, B. Schmittmann and R. K. P. Zia}
\affiliation{Center for Stochastic Processes in Science and Engineering, and
Department of Physics, Virginia Tech, Blacksburg, VA 24061-0435, USA.}
\email{jjdong@vt.edu}
\date{May 1, 2006}

\begin{abstract}
In the process of translation, ribosomes read the genetic code on an mRNA
and assemble the corresponding polypeptide chain. The ribosomes perform
discrete directed motion which is well modeled by a totally asymmetric
simple exclusion process (TASEP) with open boundaries. Using Monte Carlo
simulations and a simple mean-field theory, we discuss the effect of one or
two ``bottlenecks'' (i.e., slow codons) on the production rate of the final
protein. Confirming and extending previous work by Chou and Lakatos, we find
that the location and spacing of the slow codons can affect the production
rate quite dramatically. In particular, we observe a novel ``edge'' effect,
i.e., an interaction of a single slow codon with the system boundary. We
focus in detail on ribosome density profiles and provide a simple
explanation for the length scale which controls the range of
these interactions.
\end{abstract}

\pacs{05.70.Ln, 64.90.+b, 87.14.Gg}



\maketitle

\section{Introduction\protect\newline }

Models and methods from nonequilibrium statistical physics find natural
applications in many biological systems, and problems from biology have
inspired many nonequilibrium models. A particularly famous case is
translation, or protein synthesis, which motivated \cite{MG} one of the most
paradigmatic nonequilibrium models, namely, the totally asymmetric exclusion
process (TASEP) \cite{S1993,DEHP,Derrida92,Derrida,Schutz}. In this article, we
revisit this venerable problem and ask some new questions, related to the
phenomenon of codon bias \cite{Solomovici,Stenstrom,TomChou,Chou9, Chou10}.
To set the scene, we give a very brief, and necessarily concise, description
of how a protein is produced from a given gene, or to be more exactly, how
the genetic information stored in its associated mRNA is translated into a
growing polypeptide chain. Each mRNA molecule has two distinct ends, 
conventionally labelled as 3' and 5'.
In the first step (initiation), a ribosome binds
to the 5' end of an mRNA. With the help of several initiation factors, the
ribosome scans the mRNA until it encounters a start codon (usually AUG),
which sets the stage for protein synthesis. This can be a complicated
process, since the initiation efficiency depends on many factors such as
growth factors, infections by viruses, temperature and nucleotides
surrounding the start codon \cite{cell}. Then elongation drives translation
forward, i.e., the ribosome moves codon by codon along this mRNA template
until it reaches the stop codon, which terminates the translation process in
the presence of a release factor. At each codon, the aminoacyl-tRNA(aa-tRNA)
with the associated anticodon binds to the ribosome, and adds the
corresponding amino acid to the growing polypeptide chain. Again, this is a
complex multi-step process. At termination, the completed polypeptide chain
is released, and the ribosome unbinds from the 3' end of the mRNA and dissociates.
Typically, at any time, several ribosomes are bound to the mRNA, and several
protein-synthesis processes take place simultaneously. However, the
ribosomes cannot overlap or overtake one another. The released polypeptide
chain still needs to fold properly in order to function in a certain cell.
But the focus of our research resides on the process from initiation to
termination.

To model this sequence of events in a highly simplified fashion, we start
from a totally asymmetric simple exclusion process (TASEP), defined on a
one-dimensional (1D) lattice with open boundaries, occupied by particles and
holes. The particles jump to the right with a site-dependent rate, provided
the destination site is empty. They enter (exit) the lattice at the left
(right), with a given entrance (exit) rate. Each site on the lattice
represents a codon on the mRNA and the particles model the ribosomes;
injection, hopping, and drainage are associated respectively with
initiation, elongation, and termination in biological terms. The elongation
rates are commonly modeled in terms of generally accepted concentrations
(``abundances'') for the associated aa-tRNAs. While there are 64 codons,
there are only about 60 anticodons \cite{Neidhardt} to associate with
different aa-tRNAs, leading to at most 60 different hopping rates for the
whole gene (which typically contains hundreds to thousands of codons,
i.e., sites). Moreover, there are only 20 amino acids, so that certain
codons can be replaced by others (so-called synonymous codons) \emph{without}
modifying the final protein product. In the biological application, an
important observable is the protein production rate, given directly by the
particle current. Clearly, the protein production rate is one essential
factor for determining the gene expression level; another one is of course
simply the associated mRNA concentration.

Clearly, this simple model falls short of the biological system in several
significant aspects. One is that the ribosome ``covers'' several codons \cite
{Heinrich,Kang}, as opposed to a particle occupying only a single site. A
TASEP with extended objects was first investigated by \cite{MG}, and more
recently in \cite{LBSZia}. 
Another is that we model multi-step processes, such as initiation and
elongation, in terms of just one rate. Therefore, we should not expect our
findings to be fully quantitative. However, we believe studying
appropriate simple models can provide crucial insights into universal
properties that lead to useful predictions.

In many biological or medical investigations, it is desirable to maximize or
minimize the production of a particular protein. In the following, we focus
on maximizing (``optimizing'') its production, but our analysis can easily
be applied to the opposite goal. In other words, we have to identify the
rate-limiting step, and attempt to modify it. Here, we assume that
the \emph{availability} of the required aa-tRNA, rather than some \emph{%
internal} reaction rate, controls the time scale on which the ribosome moves
from one codon to the next. It is then quite intuitive, and will be shown
for the model below, that the aa-tRNA with the lowest concentration controls
the protein production rate. In order to enhance production, we can
either over-express the rare aa-tRNA, or attempt to swap the associated
codons for synonymous ones which employ a more abundant aa-tRNA. Here, we
explore some aspects of the second mechanism; results for the first will be reported
elsewhere \cite{Dong-un}. Codons associated with rare (abundant) aa-tRNAs
will be termed ``slow'' (``fast'').

Obviously, swapping \emph{all} slow codons for faster ones maximizes the
production rate, but will require a significant investment of laboratory
effort. It is natural to inquire if a slightly less than maximal current
can be attained with a much smaller amount of laboratory work. In other
words, is it possible to achieve a significant (if not maximal) enhancement
of the production rate by replacing only a \emph{small} number of carefully 
\emph{targeted} codons? A naive and intuitive approach would be to remove
the \emph{slowest} codons. But is the elongation rate the only factor?
Or do the locations and spacings of the slow codons also play a role? Unless
one is guided by some mathematical insight into how slow codons affect the
protein production rate, selecting the ``right codon to replace'' will be a
haphazard process of trial and error.

In this article, we attempt to provide some guidance for this selection
process, by considering a highly simplified scenario. Neglecting almost all
of the inhomogeneity of the genetic sequence, we focus on a simple
``designer gene'', consisting of many repeats of the same codon, except at
one or two locations. At these defect sites, we insert a single slow codon.
By varying the elongation rate of these special codons, as well as their
locations and spacing, we can study their effect on the protein production
rate of such a simple gene.

A closely related question, namely, how to identify the rate-limiting step
of the protein production process, was already considered in \cite{TomChou}.
Chou and Lakatos (referred to as CL in the following) placed clusters of
slow codons into an ordinary TASEP and varied their locations and spacings.
They found that a single defect lowers the production rate significantly,
and that a small number of slow codons, spaced closely together, can lower
the current by an additional factor of $2$ or more. The latter observation
is interpreted as an effective \emph{interaction} of slow sites with one
another. Our results confirm and extend their findings. In particular, we
present more precise data for a single slow site and discover that there is
indeed an ``edge effect'', i.e., an interaction of the slow site with the
system boundary, so that the particle current does depend on the position of
the slow site. This phenomenon was not noted by CL, due to the larger error
bars on their data. Also, we focus in more detail on ribosome density
profiles and provide a simple explanation for the length scale
which controls the \emph{range} of these interactions.

To set the stage for this investigation, we briefly review the analytical
results for the steady state of TASEP, as well as an earlier relevant study 
\cite{Kolo}. For a TASEP with open boundaries and homogeneous (bulk) hopping
rate $\gamma $, the steady state current is determined by the parameters $%
\alpha $ (entrance rate) and $\beta $ (exit rate), both expressed in units
of $\gamma $. The lattice size is denoted by $N$, but plays no role in the
thermodynamic limit ($N\rightarrow \infty $). The exact solutions for this
model are discussed in \cite{S1993,DEHP,Derrida92,Derrida,Schutz}. Three phases
are found: a low-density (L) phase for $\alpha <\beta $, $\alpha <1/2$, with
current $\alpha (1-\alpha )$; a high-density (H)\ phase for $\beta <\alpha $%
, $\beta <1/2$, with current $\beta (1-\beta )$; and a maximal-current (M)\
phase for $\alpha >1/2$, $\beta >1/2$, with current $1/4$. Turning to
systems with a single slow site at position $k$, \cite{Kolo} considered a
rather restricted case in which the defect site, with jump rate $q$, is
located at the \emph{center} of the lattice ($k=N/2$). For large lattices ($%
N\gg 1$), an approximate stationary solution can be found by dividing the
lattice into two separate sublattices, connected by the ``defect'' bond ($%
k,k+1$). The rate $q$ across this bond, along with the average occupancies
at sites $k$ and $k+1$, then controls the\emph{\ effective} exit rate, $%
\beta _{eff}$, from the left, and entry rate, $\alpha _{eff}$, into the
right, sublattice. Focusing on the $N\rightarrow \infty $ limit, and using
exact results for the usual TASEP, combined with a mean-field approximation
for the current through the defect bond, $J_q(k)$, the resulting phase
diagram can be determined \cite{Kolo}. For $q<1$, the phases remain
unchanged but the \emph{phase boundaries} in the phase diagram shift.
One finds 
\begin{equation}
\alpha _{eff}=\beta _{eff}\equiv q_{eff}=\frac q{1+q}  \label{q-eff}
\end{equation}
leading to the conditions $\alpha >\beta $, $\beta <q_{eff}$ for the H phase
with current $\beta (1-\beta )$; $\beta >\alpha $, $\alpha <q_{eff}$ for the
L phase with current $\alpha (1-\alpha )$; and finally, $\alpha ,\beta
>q_{eff}$ for the M phase with current 
\begin{equation}
J_q(\infty )=\frac q{(1+q)^2}\quad .  \label{kolo-J}
\end{equation}
The argument ($\infty $) reminds us that this result is only valid if $%
N,k\gg 1$. We note, for completeness, that this simple mean-field theory
can be systematically improved by considering correlations in a larger
(but still finite) neighborhood of the slow site \cite{TomChou}.  

Since we are also interested in having two defect sites (or, more precisely,
bonds) in the system, it is natural to generalize this \emph{mean-field}
approach to three
coupled sublattices, with the same (stationary) current flowing through each
of them \cite{Dong-un}. To keep the number of parameters small, we restrict
the discussion to $\alpha =\beta =1$. The two slow sites are placed at
locations $k_{1}$ and $k_{2}$, separated by a distance $d\equiv k_{2}-k_{1}$%
, and have the same rate, $q$. We find that \emph{all} effective exit and
entry rates, for each of the three sublattices, are equal, again given by $%
q_{eff}$, Eq. (\ref{q-eff}). As a result, the left sublattice is in an H
phase, the right is in an L phase, and the central sublattice, characterized
by $\alpha _{eff}=\beta _{eff}$, displays a shock, reflecting the \emph{%
coexistence} of H and L phases. In the ordinary TASEP, such shocks are found
on the coexistence line $\alpha =\beta $. Their width is microscopic \cite
{Janowsky,JandL}, i.e., the density changes from the L value ($\alpha $) to
its H counterpart ($1-\alpha $) over, typically, a few lattice spacings.
Moreover, they diffuse freely between the boundaries, so that a
configurational average results in a linear density profile. Returning to
our system with two slow bonds, these results provide us with the associated
(asymptotic)\ current, 
\begin{equation}
J_{q}(\infty )=\frac{q}{(1+q)^{2}}\quad  \label{2-slow-J}
\end{equation}
provided the two slow sites are well separated. Comparing Eqs.(\ref{kolo-J})
and (\ref{2-slow-J}), we recognize that the second slow site has no further
effect on the current. This statement is easily generalized to having two
slow sites different rates $q_{1}\neq q_{2}$; in this case, the smaller rate
(i.e., $\min \{q_{1},q_{2}\}$) sets the current through the system.

To put our work in context, let us also note that a TASEP with quenched
random rates on the \emph{entire }lattice, followed by an average over the
disorder distribution, was investigated by \cite{Harris}. \emph{Localized}
inhomogeneities, at the multicritical point $\alpha =\beta =1/2$, were
considered by \cite{HadN}. Finally, \cite{LBS} extended the work of \cite
{Kolo} to extended objects. 

This paper is organized as follows. In the next section, we present our
mathematical model and some technical details of the simulations. We then
present our data and discuss the effect of defect location and spacing on
the particle current. Finally, we turn to the implications for protein
production rates and conclude with some simple qualitative predictions.

\section{Model and Methods\protect\newline
}

We use a one-dimensional (1D) lattice of $N$ sites as a template. The
microconfiguration of the lattice can be described in terms of \emph{site
occupancies}, $n_{i}$, where the index $i=1,...,N$ is a site label. Each
site, initially chosen to be empty, is allowed to be occupied by a single
particle ($n_{i}=1$) or left empty ($n_{i}=0$). Particles enter at the left
end, jump to the neighboring site on the right provided it is empty, and
finally exit from the right end. In our random sequential updating scheme,
we select a site at random and update it, if possible, according to the
following rules:

\begin{itemize}
\item  $0\rightarrow 1$ at site $1$ with rate $\alpha $,

\item  $1\rightarrow 0$ at site $N$ with rate $\beta $,

\item  $10\rightarrow 01$ at sites $(i,i+1)$ with rate $\gamma _i$.
\end{itemize}

For the usual TASEP, the bulk rate $\gamma _{i}$ is chosen to be unity, for $%
i=1,...,N-1$. To study how slow codons influence the final protein
production rate, we modify $\gamma _{i}$ locally, by introducing one or two
slow sites. To introduce \emph{one} slow site at position $k$, we set $%
\gamma _{k}=q<1$, while $\gamma _{i}=1$ for $i\neq k$. We are particularly
interested in the relative change in the current, $J_{q}(k)$, as the
location, $k$, of the slow codon is varied. 
\begin{figure}[tbp]
\includegraphics[height=0.8in,width=3.2in]{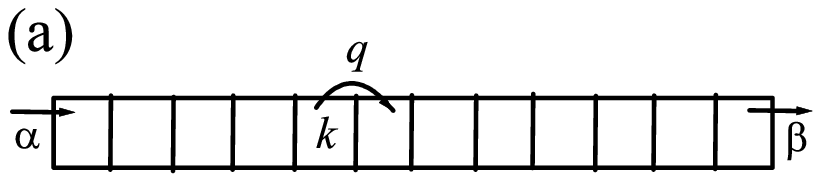} %
\includegraphics[height=0.8in,width=3.2in]{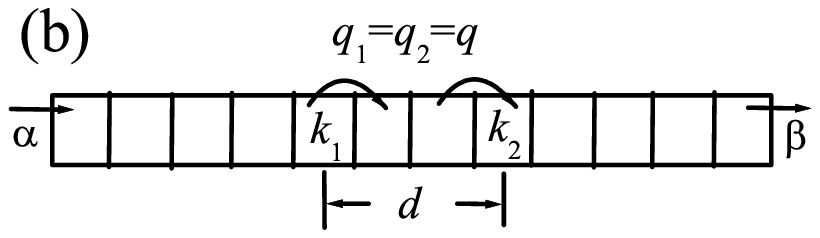}
\caption{Sketch of an ordinary TASEP with one slow site at $k$ with rate $%
q<1 $.}
\label{fig:one-slow}
\end{figure}

To introduce \emph{two} slow sites, at locations $k_{1}$ and $k_{2}$ with
separation $d\equiv (k_{2}-k_{1})$, we reduce both local rates, $\gamma
_{k_{1}}$ and $\gamma _{k_{2}}$, to $q<1$. \ Keeping in mind that slow
codons might be clustered closely together, we study the associated current,
denoted by $J_{q}(d)$, for a range of $q$ and $d$.

In our simulations, we keep a list of all occupied sites plus a single
``virtual site'' $i=0$, which is always occupied and accounts for attempted
particle entries into the system. To achieve the most efficient updating and
to reduce the number of parameters in the system, we set $\alpha =\beta
=\gamma _{i}=1$ except at one or two ``slow'' sites. At the beginning of
each Monte Carlo step (MCS), we first count the number of particles ($M$) on
the lattice. Then, we choose one of the $M+1$ list entries at random and
attempt to update it. One MCS is completed after $M+1$ update attempts have
been made. As a result, all particles on the lattice and a new particle
have, on average, experienced one update attempt. Typically, $5\times 10^{6}$
MCS are discarded to ensure that the system has reached the steady state.
Results are obtained by averaging over least $5\times 10^{4}$ measurements,
separated by $100$ MCS in order to avoid correlations. Such steady state
averages will be denoted by $\left\langle ...\right\rangle $. The system
size $N$ ranges from $200$ to $1000$, with most data taken for $N=1000$.

We monitor several observables to characterize the steady state of the
system. First, we measure the average particle current $J$, defined as the
average number of particles entering the system per unit time. By the very
definition of ``steady state'', this current is uniform throughout the
system, and could equally well be measured across any bond, or at the exit
point. In the biological system, this current corresponds to the protein
production rate. We also accumulated local density profiles, $\rho
_{i}\equiv \left\langle n_{i}\right\rangle $, to understand how they are
affected by the presence of slow sites. The overall density, $\rho \equiv 
\frac{1}{N}\left\langle \sum_{1}^{N}n_{i}\right\rangle $, follows naturally
from these profiles.

\section{Monte Carlo Simulation results\protect\newline
}

\begin{figure}[tbp]
\hfill
\begin{minipage}[t]{0.5\textwidth}
 \begin{center} 
\includegraphics[height=6cm,width=7cm]{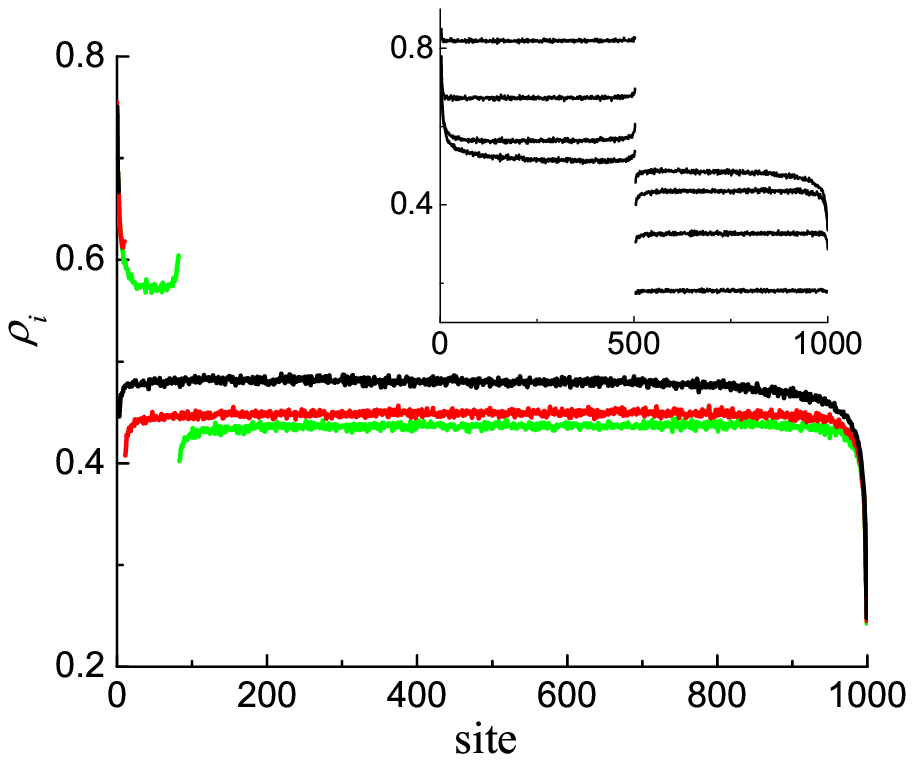} 
\vspace{-0.6cm}
\caption{Density profiles for a $N=1000$ lattice with one slow site at $k=2$
(black), $10$ (dark grey, red online) and $82$ (light grey, green online)
with $q=0.6$. Inset: Density profiles for $q=0.2$, $0.4$, $0.6$ and $0.8$
(from top to bottom on the left, and bottom to top on the right). The slow
site sits at the center ($k=500$), and $N=1000$. In all cases, the profiles
are discontinuous across the defect bond. }
\label{fig:one-profile}
 \end{center}
\end{minipage}
\hspace{0.6cm} 
\begin{minipage}[t]{0.45\textwidth}
 \begin{center}
\includegraphics[height=6cm,width=7cm]{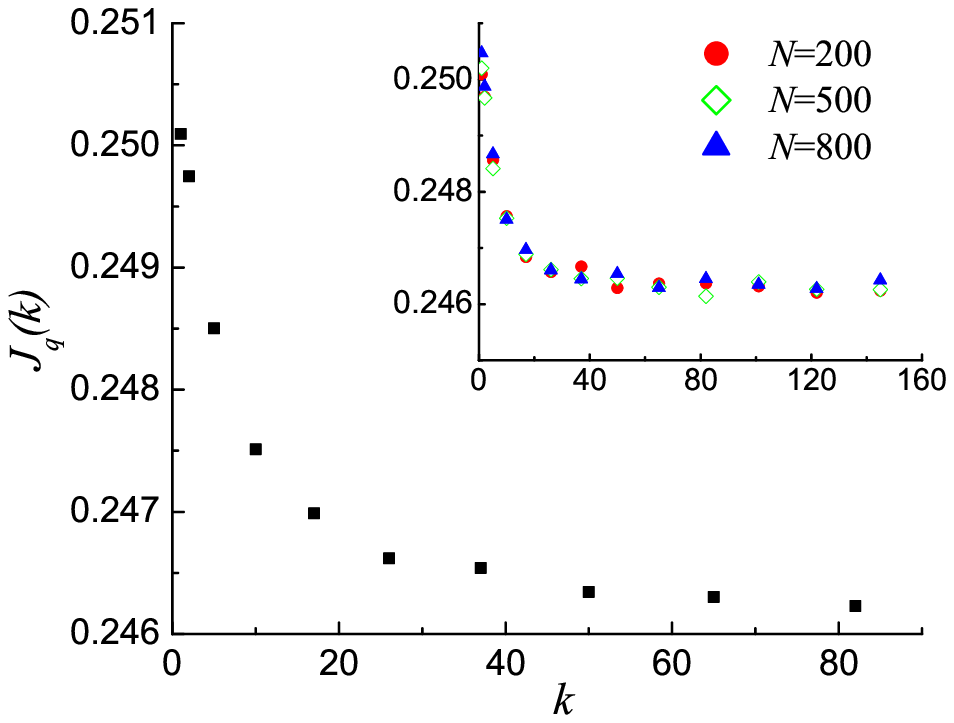} 
\vspace{-0.6cm}
\caption{$J_q(k)$ as a function of the position $k$of the slow site for $%
q=0.6$ and $N=1000$. $J_q(k)$ approaches the limit 0.2463(5) as $k
\rightarrow 500$. The inset shows that $J_q(k)$ is independent of $N$,
within statistical fluctuations. }
\label{fig:one-J}
 \end{center}
\end{minipage}
\end{figure}

We begin by placing one slow site (or defect bond) on the lattice as in Fig.~%
\ref{fig:one-slow}. Fig.~\ref{fig:one-profile} shows several density
profiles, illustrating the presence of significant non-uniformities.

\begin{figure}[b]
\includegraphics[height=6cm,width=7cm]{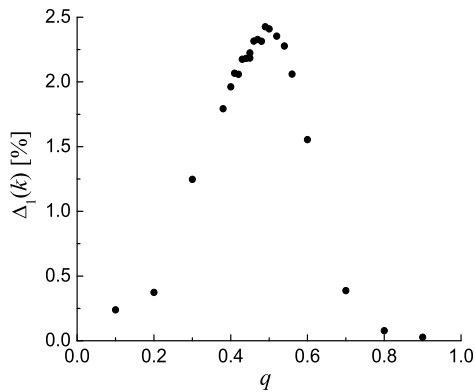} %
\caption{$\Delta _1(q)$ plotted vs. $q$, for $N$=1000.}
\label{fig:one-diff}
\end{figure}

The ``tails'', i.e., the deviations from the relatively slowly varying bulk
values, are quite noticeable in the vicinity of both the slow site and the
edges of the system. Though reminiscent of the profiles shown schematically
in \cite{HadN}, ours differ qualitatively, as a result of the loss of the $%
i\Leftrightarrow N-i-1$ symmetry ($k\neq N/2$), as well as $\alpha =\beta =1$
instead of $1/2$. Not surprisingly, there is no discernable relationship
between the profiles of the two sublattices (except in the inset). More
significantly, for our case the profiles (within each sublattice) are \emph{%
non-monotonic}, a feature that necessarily contradicts mean-field
predictions. Turning to the current, we see that, except for the smallest $q$%
's, serious deviations from Eq.~(\ref{kolo-J}) emerge. Fig.~\ref{fig:one-J},
for $q=0.6$, shows that the current increases monotonically when the slow
site is located closer and closer to the boundaries. Other choices of $q$
lead to similar behavior. Since particle-hole symmetry holds, $J_{q}(k)$ is
symmetric under inversion, $k\rightarrow N+1-k$. To quantify the $k$%
-dependence, we define a relative change in the current, 
\begin{equation}
\Delta _{1}(q)=\frac{J_{q}(1)-J_{q}(\infty )}{J_{q}(\infty )}
\end{equation}
As illustrated by Fig.~\ref{fig:one-diff}, 
the magnitude of this difference depends sensitively on $q$, reaching a
maximum at $q=0.49$ where the relative current increase is about $2.5\%$. We
refer to this phenomenon as the ``edge effect''. Since the current through
the left and the right sublattices is controlled by the bulk densities
there, our findings immediately imply that these bulk densities, denoted by $%
\rho _{bulk}$, also shift with $k$. This feature is clearly displayed in
Fig. \ref{fig:one-profile}.

Returning to Fig.~\ref{fig:one-J}, we note that significant deviations from
the limiting value, $J_{q}(\infty )$, are limited to a narrow window of $%
\delta \simeq 20$ sites near the boundaries. Thanks to charge-parity (CP)
invariance, both entry and exit edges display identical behaviors, therefore
we may restrict ourselves to, e.g., the region near the entrance. We believe
that the origin of this length scale can be traced to the presence of
exponential tails in the density profiles of the ordinary TASEP. For a
homogeneous TASEP in the H phase, with entrance and exit rates $\alpha $ and 
$\beta $, the density decays exponentially into the bulk, as $\rho _{\ell
}-\rho _{bulk}\sim \exp (-\ell /\xi )$. For $\alpha >1/2$, the decay length
becomes independent of $\alpha $ and is given by \cite{Schutz} 
\begin{equation}
\xi (\beta )=-\frac{1}{\ln \left[ 4\beta (1-\beta )\right] }  \label{H-xi}
\end{equation}
In our case, we have $\alpha =1$, while $q_{eff}=q/(1+q)$ plays the role of $%
\beta $. Thus, for the $q=0.6$ case, we find $\xi (q_{eff})\simeq 15.5$. If
the slow site is placed so close to the boundary that $k\lesssim \xi
(q_{eff})$, we should certainly expect to see deviations from Eq.~(\ref
{kolo-J}) , a formula underpinned by the assumption $k\gg 1$. In support of
our conjecture, we note that first, the observed $\delta $ is consistent
with $\xi (q_{eff})$, and second, that $\delta $, like $\xi (q_{eff})$, is 
\emph{at most} weakly dependent on the system size (cf.~the inset of Fig.~%
\ref{fig:one-J}). More detailed investigations are in progress, to settle
this issue decisively \cite{Dong-un}. In particular, if power laws such as
the ones observed by \cite{HadN} were to prevail, this picture would have to
be revised.

\begin{figure}[t]
\hfill
\begin{minipage}[t]{0.5\textwidth}
 \begin{center} 
\includegraphics[height=6cm,width=7cm]{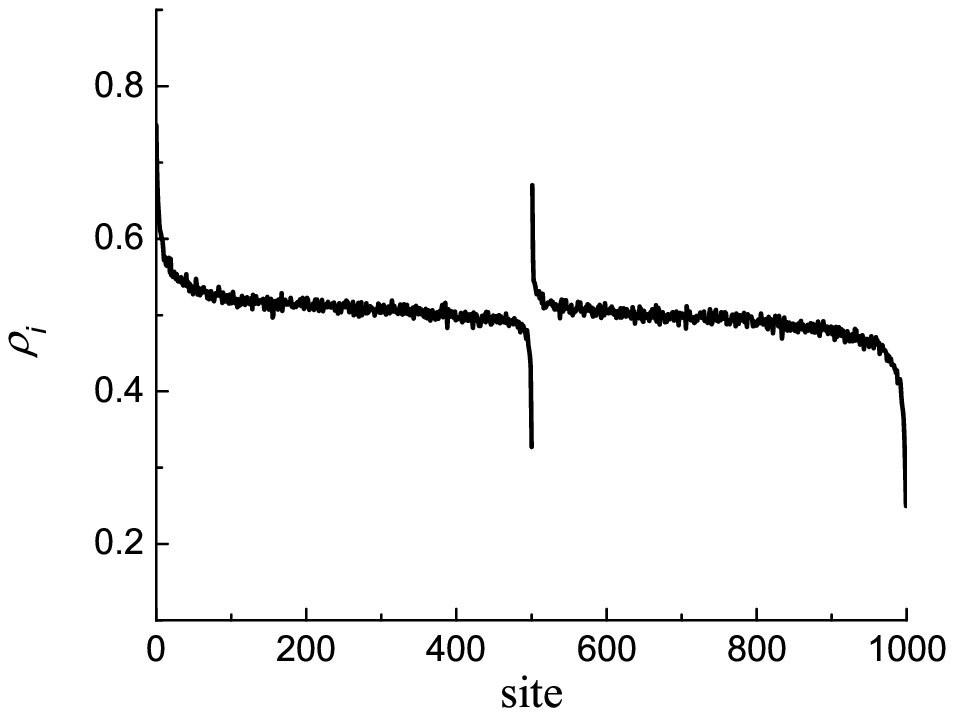} 
\vspace{-0.6cm}
\caption{Density profile for $q=\infty $, $k=500$ and $N$=1000. }
\label{fig:one-fast-profile}
 \end{center}
\end{minipage}
\hspace{0.6cm} 
\begin{minipage}[t]{0.45\textwidth}
 \begin{center}
\includegraphics[height=6cm,width=7cm]{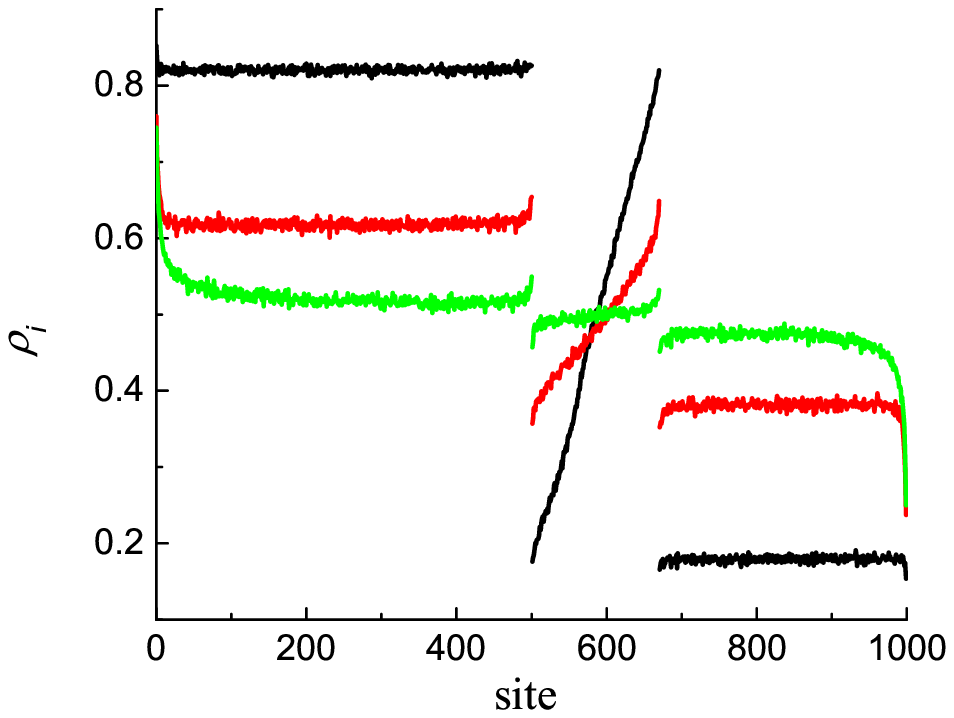} 
\vspace{-0.6cm}
\caption{Density profiles for two slow sites, both with $q=0.2$ (black), $0.5
$ (dark grey, red online), and $0.8$ (light grey, green online). The system
size is $N=1000$, and the two slow sites are located at $k_1=500$, and $%
k_2=670$, resulting in $d=170$.}
\label{fig:two-profile}
 \end{center}
\end{minipage}
\end{figure}

According to the mean-field theory described in the former section, the
presence of a defect with $q>1$ (a ``fast site''), located at the center of
the lattice, should have no noticeable effect on the current \cite{Kolo}. 
Of course, it
is not immediately apparent whether this statement remains true if the fast
site is moved closer to the system boundaries. To explore whether such a an
edge effect emerges, we consider the extreme case of $q=\infty $. Our
simulation results confirms that the current does indeed remain unchanged.
We find $J_{q}(k)=1/4+$ $O(1/N)$, consistent with the expected behavior of
the M phase. In contrast to the current, the density profiles display a
dramatic signature of the fast bond, as illustrated by Fig.~\ref
{fig:one-fast-profile}. 

If we consider the edge effect as an interaction of the slow site with the
lattice boundaries, the natural next step is to explore the interactions
between \emph{two} slow sites \cite{TomChou}. In order to avoid edge
effects, we place the two slow sites sufficiently far away from the
boundaries and vary their separation.

\begin{figure}[b]
\hfill
\begin{minipage}[t]{0.5\textwidth}
 \begin{center} 
\includegraphics[height=6cm,width=7cm]{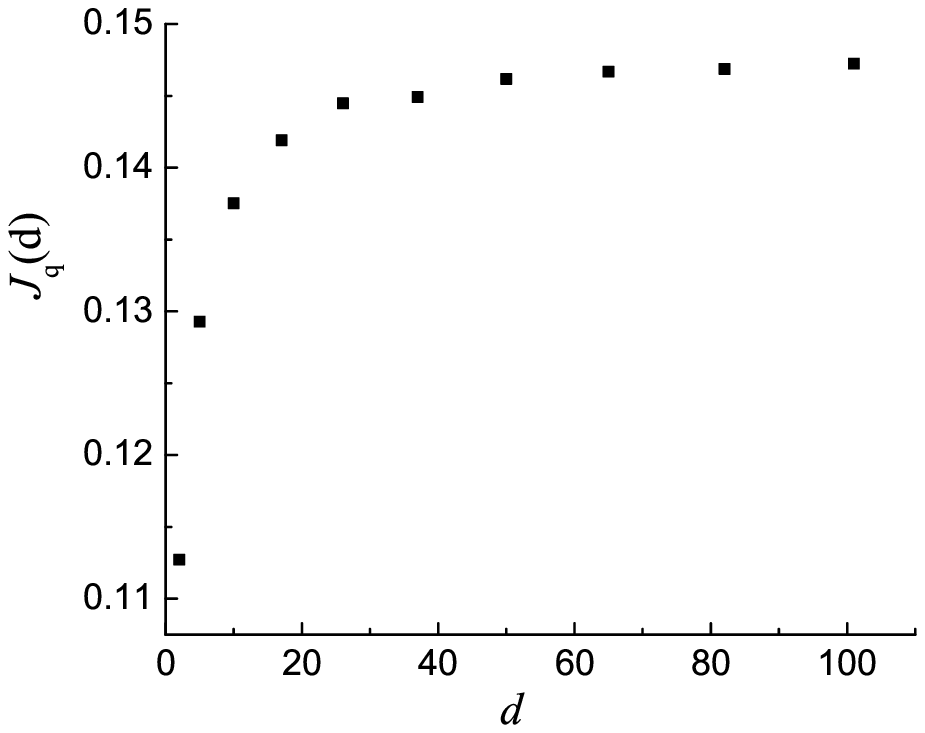}
\vspace{-0.7cm}
\caption{$J_q(d)$ for $q=0.2$ and $N=1000$, as a function of $d$. One slow
site is located at $k_1=500$, and $k_2$ is varied. }
\label{fig:two-J}
 \end{center}
\end{minipage}
\hspace{0.6cm} 
\begin{minipage}[t]{0.45\textwidth}
 \begin{center}
\includegraphics[height=6cm,width=7cm]{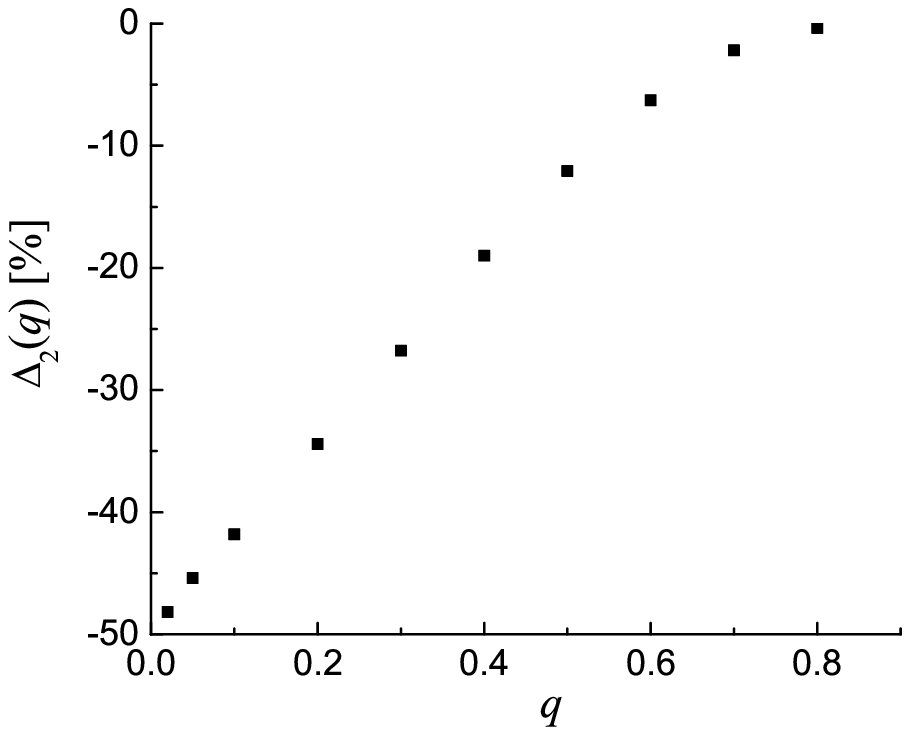}
\vspace{-0.7cm}
\caption{$\Delta _2(q)$ plotted vs. $q$, for $N$=1000. See also
Fig.~5(c) in \cite{TomChou}.}
\label{fig:two-diff}
 \end{center}
\end{minipage}
\end{figure}

Fig.~\ref{fig:two-profile} shows several typical density profiles. If $q$ is
rather small (e.g., $0.2$), CL already noted the expected linear behavior in
the central section, caused by the ``wandering shock.'' For larger $q$,
however, the center profile begins to develop distinct tails near the two
slow sites. Turning to the current, $J_q(d)$, we see from Fig.~\ref
{fig:two-J} that it is consistent with Eq.~(\ref{2-slow-J}), for $d\gtrsim 50
$, up to a finite-size correction of $O(1/N)$. In contrast, confirming the
results of CL, we observe significant deviations from Eq.~(\ref{2-slow-J}),
when $d$ is decreased. We quantify the difference by defining 
\begin{equation}
\Delta _2(q)=\frac{J_q(1)-J_q(\infty )}{J_q(\infty )}
\end{equation}
where the arguments now refer to the \emph{distance} between the two slow
sites. In contrast to $\Delta _1$, we observe that $\Delta _2$ exhibits a 
\emph{sizable} dependence on $q$, especially for small values of $q$.
Indeed, as already noted by CL, one can show that, in the limit of $%
q\rightarrow 0$ the current decreases by a factor of 2. The data in Fig.~\ref
{fig:two-diff} are clearly consistent with this conclusion. To sum up in
words, two bottlenecks near each other have a dramatic effect on the
current. Following CL, we may regard this phenomenon as an ``interaction''
between the two slow sites, inducing far more ``resistance'' when they are
close than when they are well-separated. 

Two additional comments are in order. First, we return to one of the
predictions of the mean-field theory, namely that a second slow site, spaced
far apart from its partner, should have no further effect on the current.
Our data indicate that the current for two slow sites, spaced far apart, is
systematically \emph{lower} than the current for a single slow site, but
only by a very small amount (less than $1\%$). Second, we can again attempt
to identify a length scale which controls how $J_{q}(d)$ approaches $%
J_{q}(\infty )$, as $d$ increases. Since the central section of the system
displays a shock, it is natural to ask whether the intrinsic width of the
shock sets this length scale. According to \cite{Janowsky,JandL}, this width
covers only a few lattice spacings in the \emph{periodic} TASEP with a
single defect. Here, however, it appears that the shock \emph{broadens};
preliminary data \cite{Dong-un} indicate a width of about $40$ sites for $%
q=0.2$, which is not inconsistent with Fig.~\ref{fig:two-J}. Again, this
behavior appears to be independent of the system size $N$. More work is
needed to fully explore this intriguing characteristic.

\section{Summary, Conclusions and Outlook\protect\newline }

To summarize, we study an inhomogeneous open-boundary TASEP with one or two
``slow'' sites ($q<1)$. Our key findings are as follows: For a single slow
site, there is an ``edge effect'': moving the defect closer to the boundary
enhances the current. The relative enhancement depends on $q$, but the
effect is relatively small under all model conditions (at most $2.5\%$, for $%
q=0.49$). A single fast site, on the other hand, has no effect on the
current, irrespective of its location. A much more significant effect, with
clear biological implications, emerges in the case of two slow sites. This
was already noted by CL, and we confirm their findings:\ As a function of
the separation between the two sites, the current $J_{q}(d)$ decreases
significantly, as the two sites approach each other. A quantitative measure
of this effect is the fractional reduction: $\Delta _{2}(q)$. Its dependence
on $q$ (Fig.~\ref{fig:two-diff}) is nontrivial: In the $q\rightarrow 0$
limit, the current is reduced by as much as a factor of 2.

In order to gain a better understanding of these ``interactions'' between
slow sites, and between slow sites and the system boundaries, we investigate
particle (ribosome) density profiles. Every slow site displays a clear
signature in the density profile, fully consistent with those in
previous studies \cite{Kolo,HadN}. If a defect is located at site $k$,
the profile is discontinuous between $k$ and $k+1$, and this discontinuity
is surrounded by a ``boundary layer'', or ``tails'', where the densities
deviate significantly from their bulk (asymptotic) values. In addition, the
profiles display boundary layers near the system edges, as in the ordinary
TASEP. When a defect is placed so close to a system edge or another defect
that these boundary layers begin to overlap, the particle current develops a
sensitivity to the defect-defect or defect-edge separation. In all other
cases, the current is limited by the slowest codon in the system. In this
sense, the slowest codon acts as a ``gate keeper''.

The above findings are significant in the sense that these currents are
directly linked to the protein production rate. Therefore, our results
should be directly applicable to ``designer genes'', repeating the same
codon, except at one or two locations. If the defect codons are ``fast'',
i.e., associated with a highly abundant aa-tRNA, the production rate of the
corresponding protein is insensitive to the presence of the defect codons,
but the ribosome distribution on the mRNA will display a kink. In contrast,
if the defect codons are ``slow'', i.e., associated with rare aa-tRNAs, the
protein production rate is significantly reduced. The magnitude of the
reduction depends on the locations of the slow codons. A single slow codon
near the beginning (or end) of the gene allows for a higher production rate
than a single slow codon further away, and two slow codons placed next to
one another generate a much more drastic reduction than two slow codons
spaced far apart. Preliminary studies \cite{Dong-un} indicate that our
findings remain qualitatively correct even if the particles (ribosomes)
cover more than just one site.

We can venture some even more wider-ranging predictions. Since several
different aa-tRNA (anticodons) can be associated with the same amino acid,
it is possible to produce the same protein from several different codon
sequences which will have different production rates. Given a particular
gene, we can obviously maximize the production rate of its associated
protein by systematically replacing all slow codons with synonymous, faster
ones. However, in many genes this requires a large number of substitutions
which tends to be impracticable. Instead, our findings lead us to believe
that we can pinpoint a small number (two or three) of selected substitutions
(focusing on the slowest codons, or groups of several slow codons clustered
together) which lead to nearly optimized production rates, with considerably
less effort. Preliminary data for real codon sequences lend first support to
this conjecture, and work is in progress to test these ideas more thoroughly
in silicon and in vitro \cite{Dong-un}.
\begin{acknowledgments}
We have benefitted from discussions with M. Evans, M. Ha, R. Kulkarni, P.
Kulkarni, M. den Nijs, S.-C. Park, L.B. Shaw, and B. Winkel. This work is
supported in part by the NSF through DMR-0414122 and NSF DGE-0504196. JD
also acknowledges generous support from the Virginia Tech Graduate School.%
\emph{\newline
}
\end{acknowledgments}


\begin{thebibliography}{99}
\bibitem{MG}  C. MacDonald, J. Gibbs, and A. Pipkin, Biopolymers, \textbf{6}%
, 1 (1968); C. MacDonald and J. Gibbs, Biopolymers, \textbf{7}, 707 (1969).

\bibitem{Derrida92}  B. Derrida, E. Domany, and D. Mukamel, J. Stat. Phys. 
\textbf{69}, 667 (1992).

\bibitem{DEHP}  B. Derrida, M.R. Evans, V. Hakim, and V. Pasquier, 
J. Phys. A: Math. Gen. \textbf{26}, 1493 (1993).

\bibitem{S1993}  G.M. Sch\"{u}tz and E. Domany, J. Stat. Phys. \textbf{72},
277 (1993).

\bibitem{Derrida}  B. Derrida, Phys. Rep. \textbf{301}, 65 (1998).

\bibitem{Schutz}  G.M. Sch\"{u}tz, in \textit{Phase Transition and Critical
Phenomena} edited by C. Domb and J.L. Lebowitz (Academic Press, San Diego,
2000).

\bibitem{Solomovici}  J. Solomovici, T. Lesnik and C. Reiss, J. Theor. Biol, 
\textbf{185}, 511 (1997).

\bibitem{Stenstrom}  C.M. Stenstr\"{o}m, H. Jin, L.L. Major, W.P. Tate, and
L.A. Isaksson, Gene \textbf{263}, 273 (2001).

\bibitem{TomChou}  T. Chou and G. Lakatos, Phys. Rev. Lett. \textbf{93},
198101 (2004).

\bibitem{Chou9}  M. Robinson, R. Lilley, S. Little, J.S. Emtage, G.
Yarranton, P. Stephens, A. Millican, M. Eaton, and G. Humphreys, Nucleic
Acids Res. \textbf{12}, 6663 (1984).

\bibitem{Chou10}  M.A. Sorensen, C.G. Kurland, and S. Pedersen, J. Mol.
Biol. \textbf{207}, 365 (1989).

\bibitem{cell}  B. Alberts, A. Johnson, J. Lewis, M. Raff, K. Roberts, and
P. Walter, in \textit{Molecular biology of the cell}, 4th ed. (Garland
Science, New York, NY, 2002)

\bibitem{Neidhardt}  F. Neidhardt and H. Umbarger, in \textit{Escherichia
coli and Salmonella}, 2nd ed. edited by F.C. Neidhardt (ASM Press,
Washington D.C., 1996).

\bibitem{Heinrich}  R. Heinrich and T. Rapoport, J. Theo. Biol. \textbf{86},
279 (1980).

\bibitem{Kang}  C. Kang and C. Cantor, J. Mol. Struct. \textbf{181}, 241
(1985).

\bibitem{LBSZia}  L.B. Shaw, R.K.P. Zia, and K.H. Lee, Phys. Rev. E \textbf{%
68}, 021910 (2003).

\bibitem{Dong-un}  J.J. Dong, B. Schmittmann, and R.K.P. Zia, to be
published.

\bibitem{Kolo}  A. Kolomeisky, J. Phys. A: Math. Gen. \textbf{31}, 1153
(1998).

\bibitem{Janowsky}  S. Janowsky and J. Lebowitz, Phys. Rev. A \textbf{45},
618 (1992).

\bibitem{JandL}  S. Janowsky and J. Lebowitz, J. Stat. Phys. \textbf{77}, 35
(1994).

\bibitem{Harris}  R.J. Harris and R.B. Stinchcombe, Phys. Rev. E \textbf{70}%
, 016108 (2004).

\bibitem{HadN}  M. Ha, J. Timonen, and M. den Nijs, Phys. Rev. E \textbf{68}%
, 056122 (2003). For more details, see also M. Ha, PhD thesis, University of
Washington, 2003.

\bibitem{LBS}  L.B. Shaw, A.B. Kolomeisky and K.H. Lee, J. Phys. A: Math.
Gen. \textbf{37}, 2105 (2004).
\end{thebibliography}
\end{document}